\documentclass{aip-cp}

\usepackage[numbers]{natbib}
\usepackage{rotating}
\usepackage{graphicx}

\begin{document}

\title{New Statistical PDFs: Predictions and Tests up to LHC Energies} 

\author[aff1]{Jacques Soffer\corref{cor1}}
\author[aff2]{Claude Bourrely}

\affil[aff1]{Temple University, Department of Physics, 1925 N. 12th St., Philadelphia, PA 19122, USA}
\affil[aff2]{Aix Marseille Univ,  Univ Toulon, CNRS, CPT, Marseille, France}
\corresp[cor1]{jacques.soffer@gmail.com}

\maketitle

\begin{abstract}
The quantum statistical parton distributions approach proposed more than one decade ago is 
revisited by considering a larger set of recent and accurate Deep Inelastic Scattering experimental 
results. It enables us to improve the description of the data by means of a new determination of the
 parton distributions. This global next-to-leading order QCD analysis leads to a good description of
 several structure functions, involving unpolarized parton distributions and helicity distributions, in a 
 broad range of $x$ and $Q^2$ and in terms of a rather small number of free parameters. There are 
 several challenging issues, in particular the behavior of $\bar d(x) / \bar u(x)$ at large $x$, a possible 
 large positive gluon helicity distribution, etc.. The predictions of this theoretical approach will be tested 
 for single-jet production and charge asymmetry in $W^{\pm}$ production in $\bar p p$ and $p p$ collisions 
 up to LHC energies, using recent data and also for forthcoming experimental results.

\end{abstract}

\section{INTRODUCTION}
Deep Inelastic Scattering (DIS) of leptons and nucleons is indeed our main
source of information to study the internal nucleon structure in terms of
parton distributions. Several years ago a new set of parton distribution
functions (PDFs) was constructed in the framework of a statistical approach 
of the nucleon \cite{bbs1}. For quarks (antiquarks), the building blocks 
are the helicity dependent distributions $q^{\pm}(x)$ ($\bar q^{\pm}(x)$). 
This allows to describe simultaneously the unpolarized distributions 
$q(x)= q^{+}(x)+q^{-}(x)$ and the helicity distributions $\Delta q(x) = q^{+}(x)-q^{-}(x)$ 
(similarly for antiquarks). At the initial energy scale $Q_0^2$, these distributions 
are given by the sum of two terms, a quasi Fermi-Dirac function and a helicity 
independent diffractive contribution. The flavor asymmetry for the light sea, 
{\it i.e.} $\bar d (x) >\bar u (x)$, observed in the data is built in. This is simply 
understood in terms of the Pauli exclusion principle, based on the fact that the proton contains
two up-quarks and only one down-quark. We predict that $\bar d(x) /\bar u(x)$ 
must remain above one for all $x$ values and this is a real challenge for our approach, 
in particular in the large $x$ region which is under experimental investigation at the moment. 
The flattening out of the ratio $d(x)/u(x)$ in the high $x$ region, predicted by the 
statistical approach, is another interesting challenge worth mentioning. The chiral 
properties of QCD lead to strong relations between $q(x)$ and $\bar q (x)$.
For example, it is found that the well established result $\Delta u (x)>0 $\
implies $\Delta \bar u (x)>0$ and similarly $\Delta d (x)<0$ leads to $\Delta \bar d (x)<0$. 
This earlier prediction was confirmed by recent polarized DIS data and it was also 
demonstrated that the magnitude predicted by the statistical approach is compatible 
with recent BNL-RHIC data on $W^{\pm}$ production \cite{bbsW}. In addition we found the
approximate equality of the flavor asymmetries, namely $\bar d(x) - \bar u(x)
\sim \Delta \bar u(x) - \Delta \bar d(x)$. Concerning the gluon, the unpolarized 
distribution $G(x,Q_0^2)$ is given in terms of a quasi Bose-Einstein function, 
with only {\it one free parameter}. The new analysis of a larger set of recent 
accurate DIS data leads to the emergence of a large positive gluon helicity
distribution, giving a significant contribution to the proton spin, a major
point which was emphasized in a recent letter \cite{bs14}.\\ 
It is crucial to note that the quantum-statistical approach differs from the usual 
global parton fitting methodology for the following reasons:\\
i) It incorporates physical principles to reduce the number of free parameters which 
have a physical interpretation.\\
ii) It has very specific predictions, so far confirmed by the data.\\
iii) It is an attempt to reach a more physical picture on our knowledge of the nucleon 
structure, the ultimate goal being to solve the problem of confinement.\\
iv) Treating simultaneously unpolarized distributions and helicity distributions, a unique 
siuation in the literature, has the advantage to give access to a vast set of experimental data, 
in particular up to LHC energies.\\

\section{REVIEW OF THE STATISTICAL PARTON DISTRIBUTIONS}
Let us now recall the main features of the statistical approach for building up
the PDFs, as opposed to the standard polynomial type
parameterizations of the PDF, based on Regge theory at low $x$ and on counting
rules at large $x$.
The fermion distributions are given by the sum of two terms,
a quasi Fermi-Dirac function and a helicity independent diffractive
contribution, at the input energy scale $Q_0^2=1 \mbox{GeV}^2$,
\begin{equation}
xq^h(x,Q^2_0)=
\frac{A_{q}X^h_{0q}x^{b_q}}{\exp [(x-X^h_{0q})/\bar{x}]+1}+
\frac{\tilde{A}_{q}x^{\tilde{b}_{q}}}{\exp(x/\bar{x})+1}~,
\label{eq1}
\end{equation}
\begin{equation}
x\bar{q}^h(x,Q^2_0)=
\frac{{\bar A_{q}}(X^{-h}_{0q})^{-1}x^{\bar{b}_ q}}{\exp
[(x+X^{-h}_{0q})/\bar{x}]+1}+
\frac{\tilde{A}_{q}x^{\tilde{b}_{q}}}{\exp(x/\bar{x})+1}~.
\label{eq2}
\end{equation}
We note that the universal diffractive term is absent in the quark helicity distribution $\Delta q$ 
and in the quark valence contribution $q - \bar q$.\\
In Equations.~(\ref{eq1},\ref{eq2}) the multiplicative factors $X^{h}_{0q}$ and
$(X^{-h}_{0q})^{-1}$ in
the numerators of the non-diffractive parts of the $q$'s and $\bar{q}$'s
distributions, imply a modification
of the quantum statistical form, we were led to propose in order to agree with
experimental data. The presence of these multiplicative factors was justified
in our earlier attempt to generate the transverse momentum dependence (TMD)
\cite{bbs5, bbs6}.
The parameter $\bar{x}$ plays the role of a {\it universal temperature}
and $X^{\pm}_{0q}$ are the two {\it thermodynamical potentials} of the quark
$q$, with helicity $h=\pm$. They represent the fundamental characteristics of
the model. Notice the change of sign of the potentials
and helicity for the antiquarks \footnote{~At variance with statistical
mechanics where the distributions are expressed in terms of the energy, here
one uses
 $x$ which is clearly the natural variable entering in all the sum rules of the
parton model.}.\\
For a given flavor $q$ the corresponding quark and antiquark distributions
involve {\it eight} free parameters: $X^{\pm}_{0q}$, $A_q$, $\bar {A}_q$,
$\tilde {A}_q$, $b_q$, $\bar {b}_q$ and $\tilde {b}_q$. It reduces to $\it
seven$ since one of them is fixed by the valence sum rule, $\int (q(x) - \bar
{q}(x))dx = N_q$, where $N_q = 2, 1, 0 ~~\mbox{for}~~ u, d, s$, respectively.

For the light quarks $q=u,d$,  the total number of free parameters is reduced
to $\it eight$ by taking, as in Ref. \cite{bbs1}, $A_u=A_d$, $\bar {A}_u = \bar
{A}_d$, $\tilde {A}_u = \tilde {A}_d$, $b_u = b_d$, $\bar {b}_u = \bar {b}_d$
and $\tilde {b}_u = \tilde {b}_d$. For the strange quark and antiquark
distributions, the simple choice made in Ref. \cite{bbs1}
was improved in Ref. \cite{bbs2}, but here they are expressed in terms of $\it
seven$ free parameters.\\
For the gluons we consider the black-body inspired expression
\begin{equation}
xG(x,Q^2_0) = \frac{A_Gx^{b_G}}{\exp(x/\bar{x})-1}~,
\label{eq3}
\end{equation}
a quasi Bose-Einstein function, with $b_G$ being the only free parameter, since
$A_G$ is determined by the momentum sum rule.
In our earlier works \cite{bbs1,bbs4}, we were assuming that, at the input
energy scale, the helicity gluon distribution vanishes, so
\begin{equation}
x\Delta G(x,Q^2_0)=0~.
\label{eq4}
\end{equation}
However as a result of the present analysis of a much larger set of very
accurate unpolarized and polarized DIS data, we must give up this simplifying
assumption. We are now taking

\begin{equation}
 x\Delta G(x,Q^2_0) = \frac {\tilde A_G x^{\tilde b_G}}{(1+ c_G
x^{d_G})}\!\cdot\!\frac{1}{\exp(x/\bar x - 1) } \,.
 \end{equation}

To summarize the new determination of all PDFs involves a total of {\it twenty
one} free parameters: in addition to the temperature $\bar x$ and the exponent
$b_G$ of the gluon distribution, we have {\it eight} free parameters for the
light quarks $(u,d)$, {\it seven} free parameters for the strange quarks and
{\it four} free parameters for the gluon helicity distribution. These
parameters have been determined from a next-to-leading order (NLO) QCD fit of a
large set of accurate DIS data,  unpolarized and polarized structure functions \cite{bs15}.

\section{SELECTED RECENT  RESULTS}
Once determined by DIS data only, these statistical PDFs have been tested on several hadronic processes
in a rather broad energy range, up to LHC energies, and we will recall now a few examples.\\
Let us first consider single-jet production. A very satisfactory description of data from STAR at BNL RHIC, D0 and CDF at
FNAL, ALICE, ATLAS and CMS at CERN LHC was obtained, as shown in \cite{bs15}, but we  display on Figure 1 only two of these results.
\begin{figure}[h]
{\includegraphics[width=140pt]{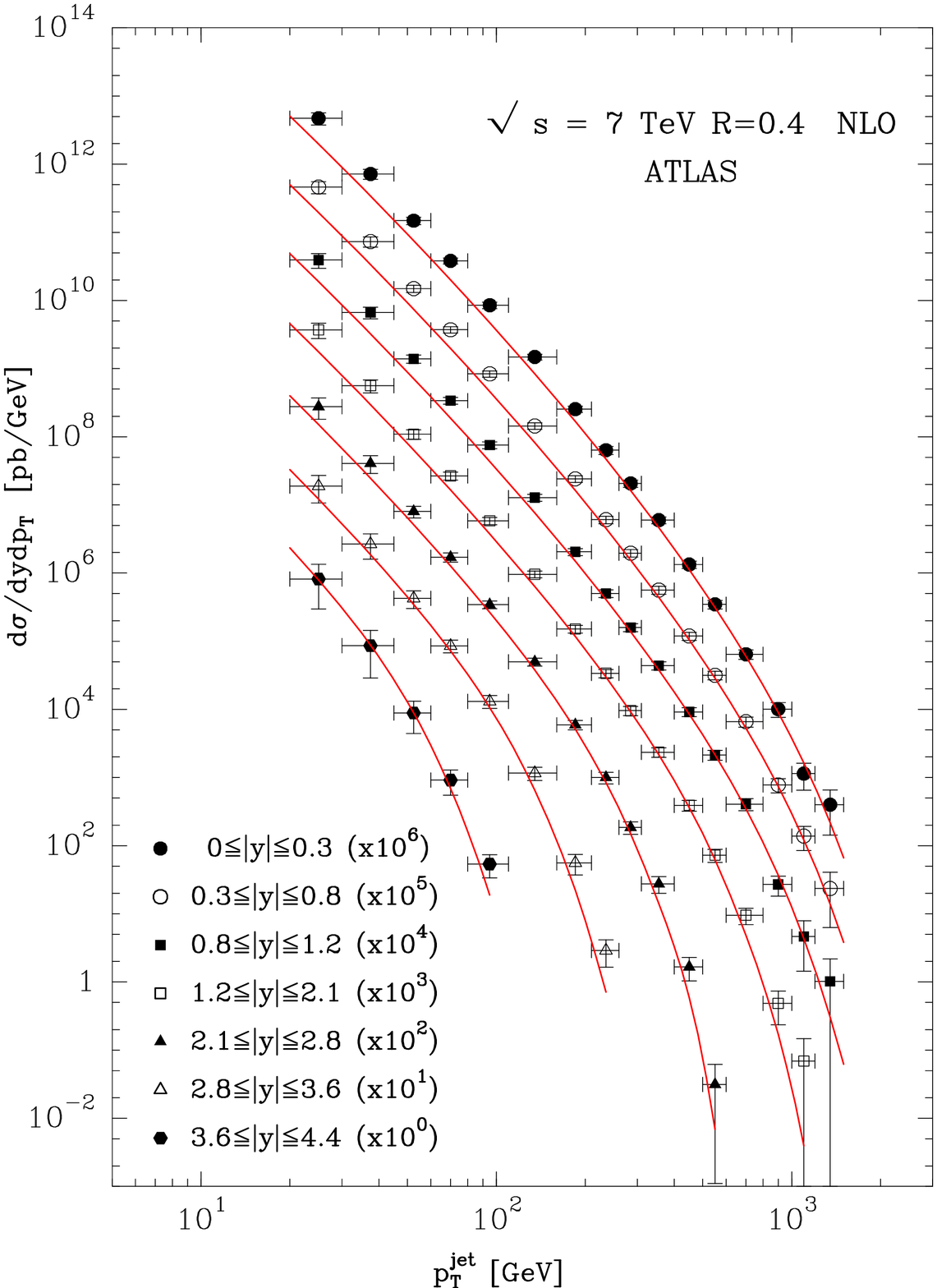}}
{\includegraphics[width=140pt]{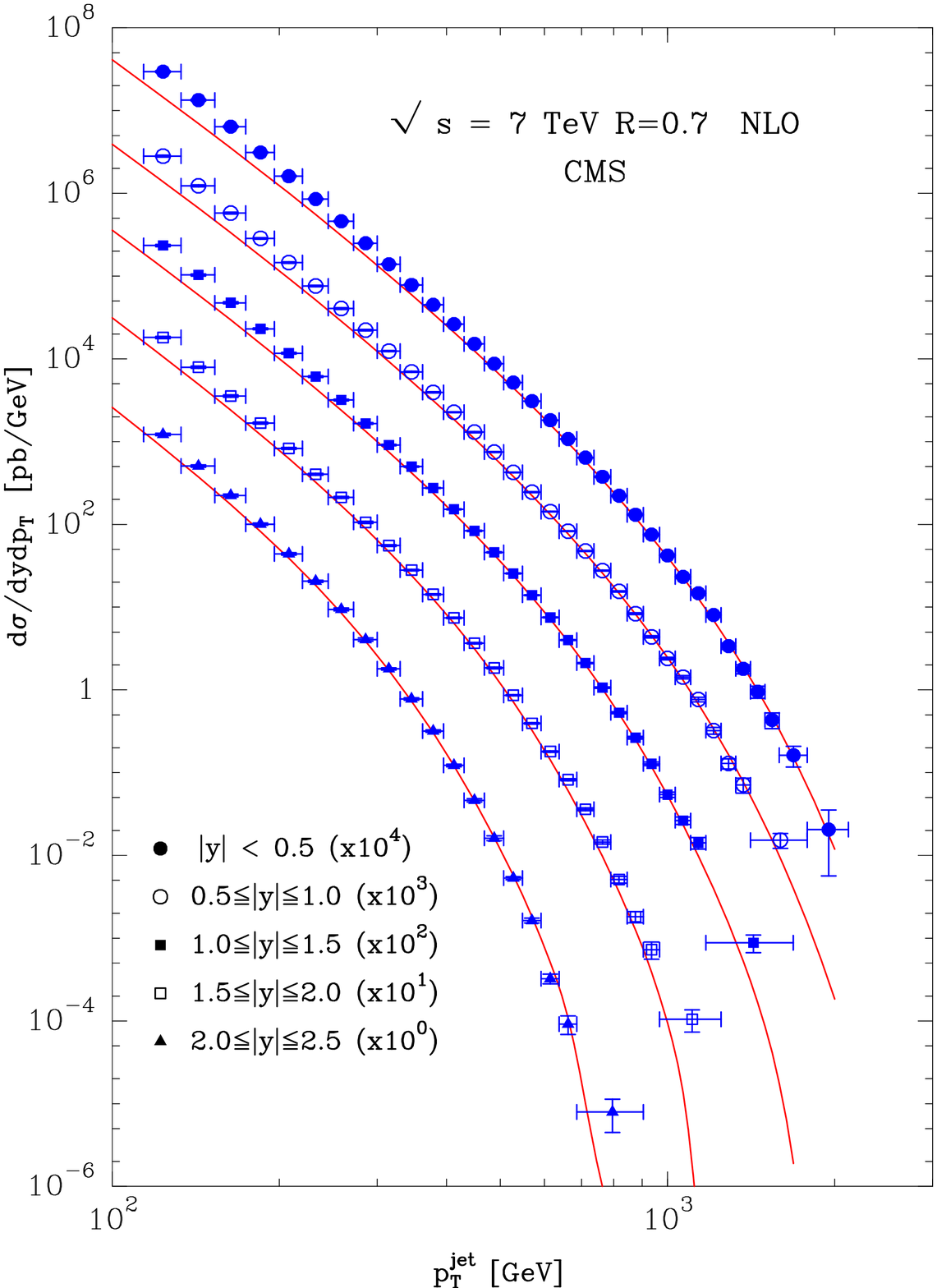}}
 \caption{
 {\it Left}:  Double-differential inclusive single-jet cross section in $pp$
collisions at $\sqrt{s}$ = 7TeV, versus $p_T^{jet}$, with jet radius parameter
R = 0.4, for different rapidity bins from ATLAS \cite{atlasjet} and the
predictions from the statistical PDFs, denoted now on as BS15.
{\it Right}: Same from CMS \cite{cms13}, with R = 0.7.}
\end{figure}
The Drell-Yan process is also a very interesting testing ground for PDFs and this was our motivation for a dedicated paper on this
important subject \cite{bbps}. Data for the $Z/\gamma^*$ production at LHC, the invariant mass distribution and the normalised rapidity distribution, are shown
on Figure 2. A comparison is made with several PDFs sets with equally remarkable good agreement.
\begin{figure}[h]
{\includegraphics[width=250pt]{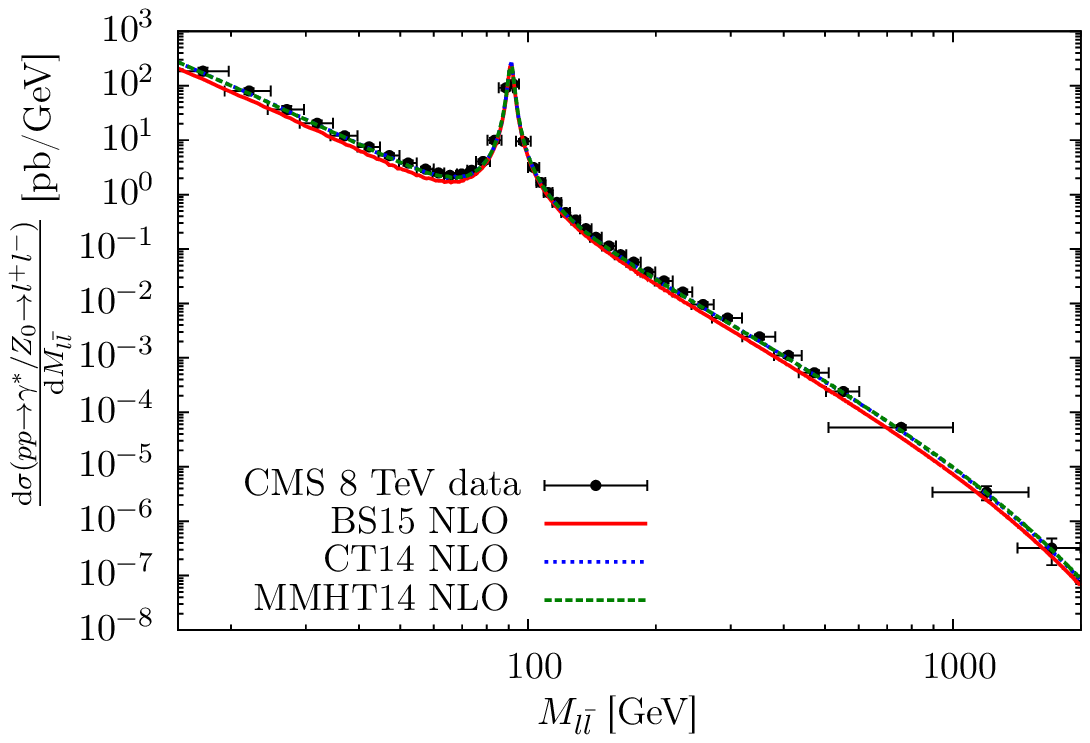}}
{\includegraphics[width=180pt]{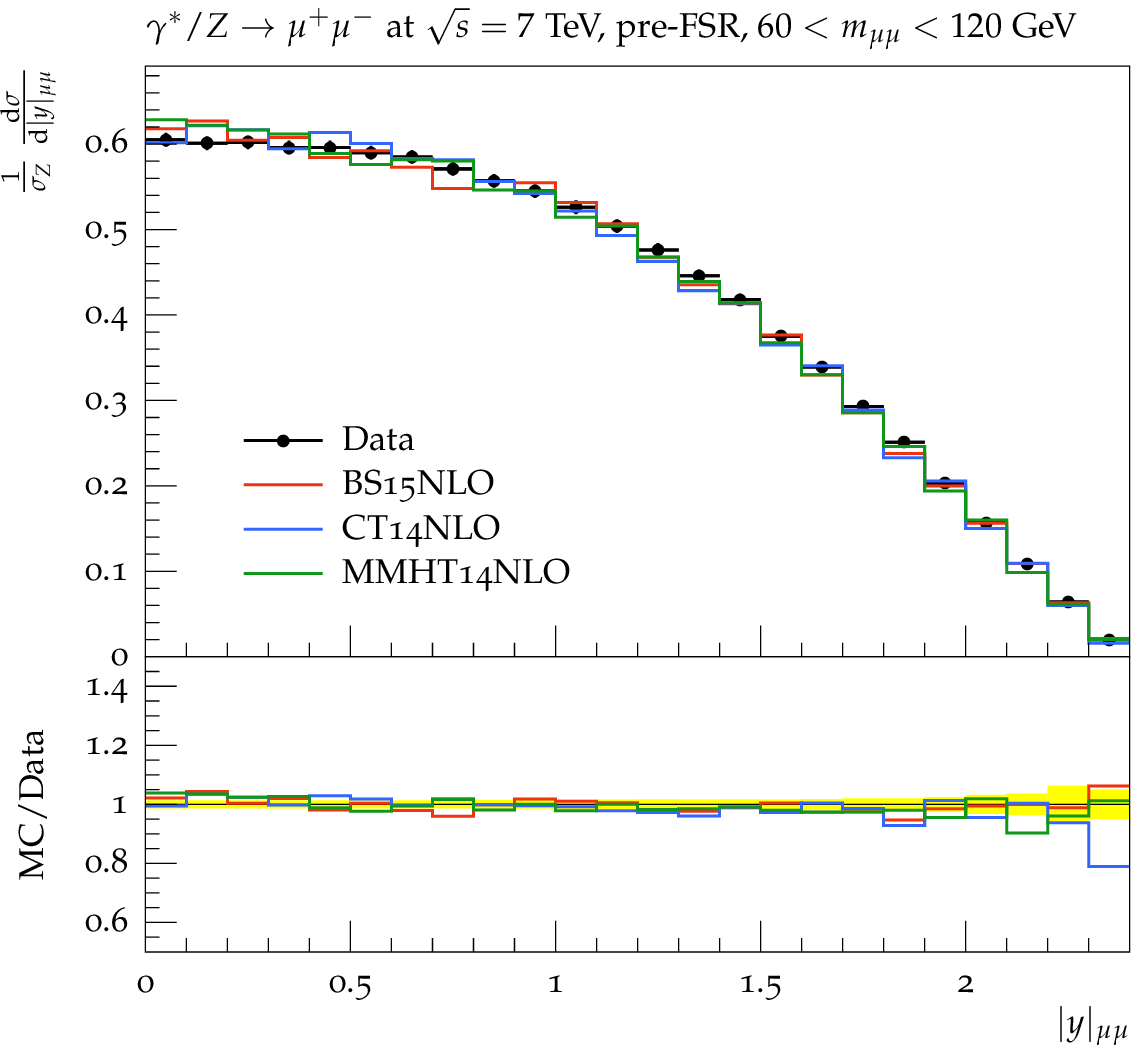}}
\caption{
 {\it Left}:  The DY differential cross section measured in the combined di-muon and di-electron channels by CMS at $\sqrt{s}$ = 8 TeV 
over the invariant mass range from 15 GeV to 2 TeV \cite{cms15} vs QCD NLO predictions 
obtained by using different PDF models.
{\it Right}: The normalized rapidity distribution measured in the di-muon channel by CMS at
$\sqrt{s}$ = 7 TeV \cite{cms13a} vs QCD NLO predictions obtained by using different
PDF models ( Taken from Ref. \cite{bbps}). }
\label{DY}
\end{figure}
\begin{figure}[h]
{\includegraphics[width=200pt]{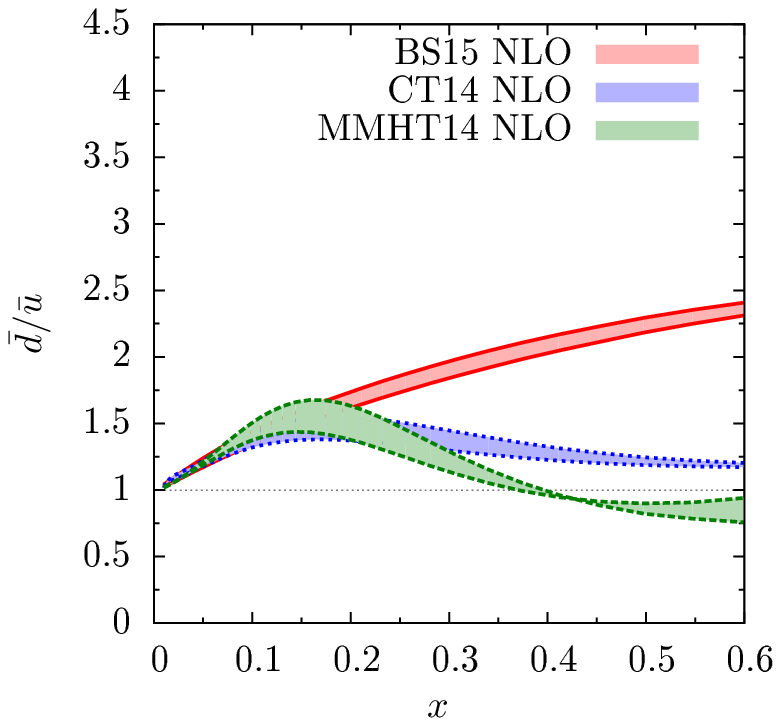}}
{\includegraphics[width=250pt]{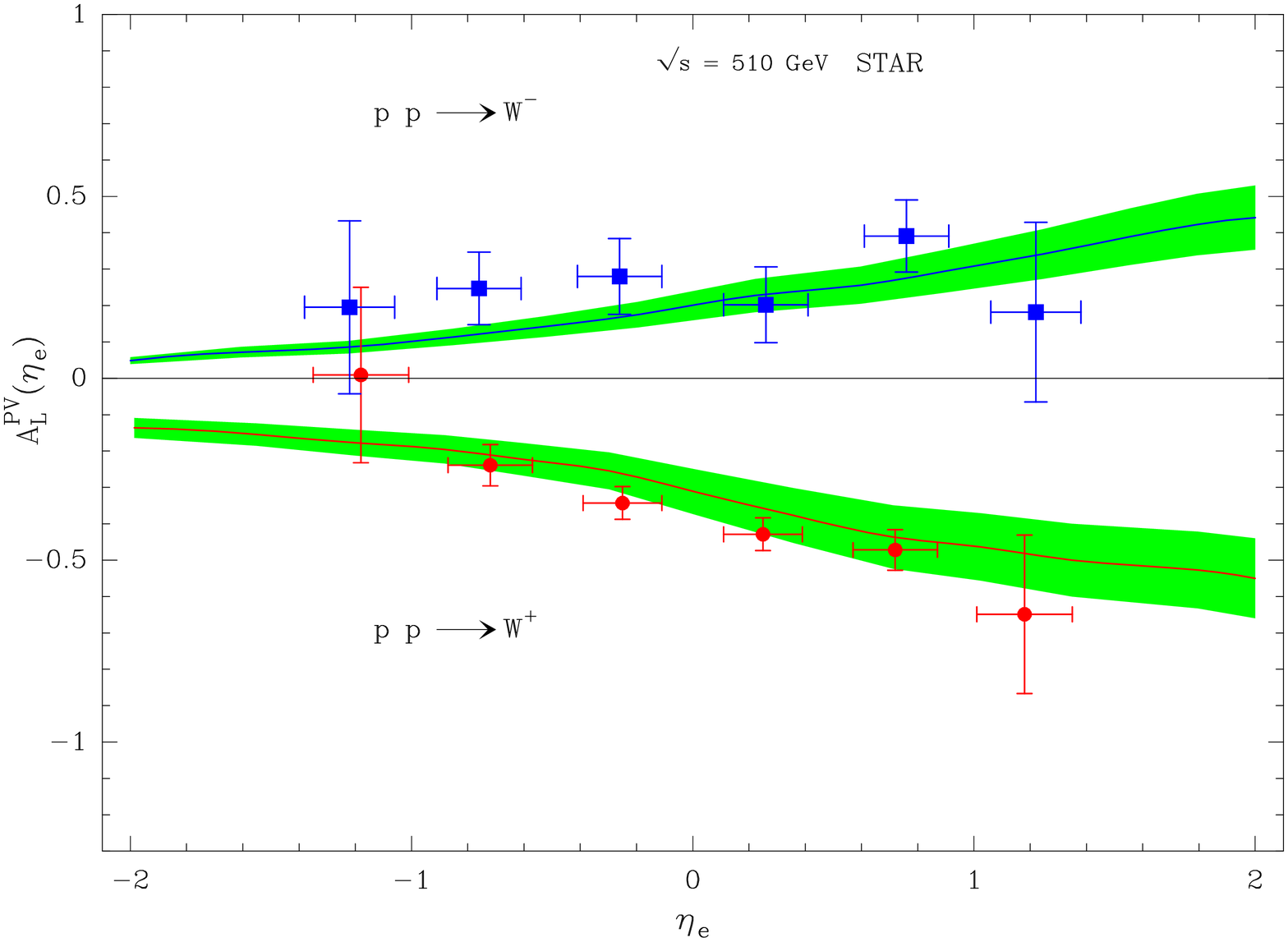}}
\caption{
 {\it Left}:  The QCD NLO predictions for the ratio of the sea quark PDFs $\bar d /\bar u$, as a function of $x$ with scale variation within $1 < Q^2 < 100 \mbox{GeV}^2$ interval,
obtained by using different PDF models  ( Taken from Ref. \cite{bbps}).
{\it Right}: The measured parity-violating helicitiy asymmetries $A_L^{PV}$ versus the pseudorapidity  of the charged-lepton $\eta_e$, from $W^{\pm}$ decay, produced at RHIC, from STAR \cite{STAR}, compared to the predictions from the statistical approach ( Taken from Ref. \cite{bbsW}). }
\label{qbar}
\end{figure}
 We will end with some considerations on the light antiquarks, which are not so well determined and deserve a special attention. First the ratio $\bar d /\bar u$ which is predicted in the statistical approach, to remains above one for all $x$ values, at variance with other PDF sets, as shown on Figure 3  ({\it Left}). This trend seems to be confirmed by the preliminary results from the SeaQuest experiment \cite{Reimer}. 
 
 Next we consider the process $\overrightarrow p p\to W^{\pm} + X \to e^{\pm} + X$, where the arrow denotes a longitudinally polarized proton and the outgoing $e^{\pm}$ have been produced by the leptonic decay of the $W^{\pm}$ boson. The helicity asymmetry is defined as  $A_L^{PV} = (d\sigma_+ - d\sigma_-)/(d\sigma_+  + d\sigma_-)$. Here $\sigma_h$ stands for the cross section where the initial proton has helicity $h$. It was measured recently at RHIC-BNL \cite{STAR} and the results are shown in Figure 3 ({\it Right}). As explained in Ref. \cite{bbsW}, the $W^-$ asymmetry is very sensitive to the sign and magnitude of $\Delta \bar u$, so this is another successful result of the statistical approach.

\end{document}